\documentclass[preprint,aps,showpacs]{revtex4}

\usepackage{bm,here}
\usepackage{graphicx}
\usepackage{amsmath,amsxtra,amsfonts,amssymb,amstext,latexsym}

\usepackage{times}
\usepackage{mathptm}

\newcommand{\p}{\partial}
\newcommand{\wh}{\widehat}
\newcommand{\wt}{\widetilde}

\newcommand{\vek}[1]{{\mathbf#1}}
\newcommand{\abs}[1]{{\left|#1\right|}}

\def\grad{\nabla}

\def\dpl{\grad_\parallel}

\def\ddpp{\grad_\perp^2}

\def\ptt#1{{\partial #1\over\partial t}}

\def\bhat{\hat\beta}
\def\muhat{\hat\mu}
\def\epss{\hat\epsilon}

\def\pxx#1{{\partial #1\over\partial x}}
\def\pzz#1{{\partial #1\over\partial z}}
\def\pyy#1{{\partial #1\over\partial y}}
\def\wcv{{\omega_B}}
\def\ppt#1{\partial #1/\partial t}
\def\ppxx#1{{\partial^2 #1\over\partial x^2}}

\def\ppyy#1{{\partial^2 #1\over\partial y^2}}

\def\vexb{\vec v_E}

\def\vedl{\vexb\cdot\grad}

\def\vor{\Omega}
\def\Psi{A_\parallel}

\begin{document}


\title{Shear Flow Generation and Energetics in  Electromagnetic Turbulence  }

\author{V.~Naulin}
\author{A. Kendl$^{\ast}$}
\author{O.~E.~Garcia}
\author{A.~H.~Nielsen}
\author{J.~Juul Rasmussen}

\affiliation{Association EURATOM-Ris{\o} National Laboratory,
OPL-128 Ris{\o}, DK-4000 Roskilde, Denmark}
\affiliation{$\ast$) University of Innsbruck, Institute for Theoretical
  Physics, Association EURATOM-\"OAW, A-6020 Innsbruck, Austria}
\date{\today}

\begin{abstract}
Zonal flows are recognised to play a crucial role for magnetised plasma
confinement. The genesis of these flows out of turbulent fluctuations
is therefore of significant interest. We investigate the relative importance
of zonal flow generation mechanisms via the Reynolds stress, Maxwell stress,
and geodesic acoustic mode (GAM) transfer in drift-Alfv\'en turbulence.  
By means of numerical computations we quantify the energy transfer into
zonal flows owing to each of these effects. 
The importance of the three driving ingredients in electrostatic and
electromagnetic turbulence for conditions relevant to the edge of fusion 
devices is revealed for a broad range of parameters. 
The Reynolds stress is found to provide a flow drive, while the
electromagnetic Maxwell stress is in the cases considered  a sink for the flow energy. 
In the limit of high plasma beta, where electromagnetic effects and Alfv\'en
dynamics are important, the Maxwell stress is found to cancel the
Reynolds stress to a high degree. The geodesic oscillations, related to
equilibrium pressure profile modifications due to poloidally asymmetric
transport, can  act as both sinks as drive terms, depending on
the parameter regime. For high beta cases the GAMs are the main
drive of the flow. This is also reflected in the frequency dependence
of the flow, showing a distinct peak at the GAM frequency in that regime. 
\end{abstract}

\pacs{
      52.25.Gj, 
      52.35.Ra, 
      52.65.Kj  
}
\maketitle

\section{Introduction} 
Since the discovery of the H-mode \cite{Wagner:Becker:Behringer:1982} in
magnetically confined plasmas a multitude of mechanisms for the generation 
of the shear flow connected to the LH-transition have been proposed. 
They include amongst others ion-orbit loss effects, neoclassical
effects, and turbulent flow generation
\cite{Connor:Wilson:2000,Hugill:2000,Terry:2000}. 
Here we focus on turbulence as a source of shear flow generation. It was
already early recognized that turbulence can lead to spontaneous
self-organization of turbulent energy into sheared poloidal flows which in
turn could reduce the transport significantly \cite{Hasegawa:Wakatani:1987}.
A conclusive computational demonstration of shear flow generation by
turbulence in realistic geometry of fusion devices, which is sufficient for
achievement of the LH-transition, has, however, not yet been achieved. 
In electrostatic turbulence the Reynolds stress is the main source
of interaction between large scale flows and small scale turbulence. 
The Reynolds stress designates the radial flux of poloidal momentum, and a
finite radial gradient of it will be an indication for a local condensation of
momentum into a poloidal flow. 
In electromagnetic turbulence an additional source of poloidal flow
generation has to be accounted for: the Maxwell stress, which arises from
parallel momentum transport along perturbed magnetic field lines. 
Measurements of the Reynolds stress and its radial variation have been
performed in several fusion devices with the purpose to identify it as a
source of sheared poloidal rotation \cite{Hidalgo:etal:2003}. 
Recently, also the Maxwell stress, respectively, magnetic fluctuations
and their cross-correlations have been measured in Reversed Field Pinch (RFP)
\cite{Antoni:IAEA:2004} and Tokamak \cite{Lu:etal:2004} configurations. 
These measurements indicate that the Maxwell stress acts as a sink for
poloidal flow energy. Finally, in the presence of toroidal magnetic field
inhomogeneity the geodesic acoustic modes (GAMs)
\cite{Winsor:Johnson:Dawson:1968} interact with the poloidal flows in the
system. In such cases the zonal flows show a residual oscillation at the GAM
frequency. 
\\
The purpose of this paper is to investigate these three different transfer
mechanisms for zonal flow generation over a wide range of parameters as neither their
strength nor their detailed (driving or damping) effect on the flows are
a priori sufficiently clear. While the Reynolds stress is most often
identified as a flow drive, there is considerable confusion about the role of
the GAMs \cite{Hallatschek:Biskamp:2001, Scott:2003}. 
The Maxwell stress is in low $\beta$ situations rather weak, but it has been
found to drain energy from the flow \cite{Wakatani:Sato:Miyato:Hamaguchi:2003}
and in high $\beta$ situations it should ideally cancel the Reynolds stress
\cite{Kim:Hahm:Diamond:2001}. 
\\
This paper is organized as follows: In the following Section \ref{sec:model}
we present the turbulence model used for the computations.  
The next Section \ref{sec:energy} is devoted to discussion of the various
transfer mechanisms of energy between turbulence and flow motion in a low and
a high beta case. We then present global scalings of the transfer terms
with collisionality and plasma beta in Sec.~\ref{sec:resglobal}.
Finally we discuss our results in the concluding section.

\section{Electromagnetic turbulence model \label{sec:model}}

We investigate the detailed balance of drive and sink terms for global
poloidal flows in a model for plasma turbulence in the edge region of magnetic
confinement devices. Considering both electrostatic and electromagnetic
effects, together with toroidal geometry and magnetic field curvature in a flux
tube model, allows us to investigate the different turbulent momentum transfer
terms responsible for flow generation. 
\\ 
The fluid equations for drift-Alfv{\'e}n turbulence in 3-dimensional flux tube
geometry result from standard ordering based upon the slowness of the dynamics
compared to the ion gyro frequency $\Omega_i=eB/M_i$ and the smallness
of the drift scale $\rho_s$ compared to the background pressure gradient scale
length $L_\perp$.  These quantities and the sound speed $c_s$ are
defined by
\begin{equation}
  \Omega_i = \frac{eB}{ M_i }, \qquad
  c_s^2 = \frac{T_e}{ M_i}, \qquad
  \rho_s = \frac{c_s}{\Omega_i}, \qquad
  L_\perp = \abs{\nabla\log p_e}^{-1},
\end{equation}
where subscripts $e,i$ refer to electrons or ions respectively,
and the temperature is given in units of energy.  Normalization is in terms of
scaled dependent variables (electrostatic potential $e\phi/T_e$, electron
density  $n/n_{00}$, parallel ion velocity $u/c_s$, parallel electric current
$J/n_{00} e c_s$). In addition the dependent quantities are scaled with
the small drift parameter $\delta=\rho_s/L_\perp$, so that mainly terms of
order one appear in the normalised set of equations. 
\\
The scale perpendicular to the magnetic field is in units of $\rho_s$; the
parallel scale is $L_\parallel=
q R$, with $R$ the toroidal major radius and $q$ the safety factor and
the closed flux surface connection length $2 \pi L_\parallel$.   
The time scale is $L_\perp/c_s$.  Further details for this system and geometry
are given in Ref.~\cite{Scott:1997:2}. 
In the present paper we use a constant background temperature for electrons,
and assume cold ions. The quantity $n_{00}$ is a normalizing density, while
$n_0(x)$ is the equilibrium plasma density having a finite gradient.
In normalized units the radial profile of the density is
$\partial_x \log n_0(x) = -1 $.  Thus $x$  serves as the radial coordinate.
Relative to the background magnetic field $\vec B$ the other perpendicular
coordinate is $y$. Finally the parallel coordinate is denoted by $z$.
\\
As singly charged ions and quasi-neutral dynamics are assumed, $n_0$
and $n$  refer to both the electron and ion density. It is important to
note that $n_0$ and $n$ equivalently describe electron density or
pressure;  we neglect temperature dynamics in this qualitative
study because of the similarity in physical character between the
electron temperature and the ``non-adiabatic'' part of the electron
density \cite{Scott:1992-1}.
The model is described by the temporal evolution of the  electrostatic
potential ($\phi$), density perturbations ($n$), parallel current ($J$), and
parallel ion velocity ($u$).
Auxiliary variables are the  vorticity ($\vor$) and the parallel component of
the magnetic vector potential ($\Psi$): 
\begin{equation}
\label{eq:eqvor}
\ptt \vor + \vedl \vor
=
{\cal{K}} \left(  n \right)
+  \nabla_{\|} J  + \mu_{\vor} \nabla_{\perp}^{2}  \vor\,,
\end{equation}
\begin{equation}
\label{eq:eqne}
\ptt n + \vedl (n_0 + n)
=
{\cal{K}} \left( n  - \phi \right)
+  \nabla_{\|} \left( J  -  u\right) +  \mu_{n}  \nabla_{\perp}^{2}  n\,,
 \end{equation}
\begin{equation}
\label{eq:eqpsi}
\ptt{} \left( \bhat\Psi + \muhat J \right) + \muhat\vedl J
=
\nabla_{\|} \left( n_0 + n - \phi \right) - C J\,,
\end{equation}
\begin{equation}
\label{eq:equi}
\epss\left(\ptt u + \vedl u\right)
=
 - \nabla_{\|} \left(n_0 + n\right) \,,
\end{equation}
with the vorticity $\vor$ and current (Ampere's law) $J$ given by
\begin{equation}
\label{eq:eqampere}
 \vor  = \nabla_{\perp}^{2} \phi\,,
\qquad\qquad
 J = - \nabla_{\perp}^{2} \Psi\,.
\end{equation}

The advective and parallel derivatives carry non-linearities entering
through $\phi$ and $\Psi$, which -- due to the description of the
geometry --
can be expressed in terms of a Poisson bracket 
\begin{equation}
{\left\{ f,g \right\} } = \pxx{f}\pyy{g}-\pyy{f}\pxx{g}
\end{equation}
in the $xy$-plane as
\begin{equation}
  \vedl = \{\phi,\cdot\}\,; \qquad\qquad
  \dpl = \pzz{} - \{\bhat\Psi,\cdot\}\,.
\end{equation}

The curvature  operator ${\cal{K}}$ is for simple circular toroidal geometry
written as 
\begin{equation}
\label{eq:eqcv}
  {\cal{K}} = - \omega_{B} \left( \sin z \pxx{}
        + \cos z \pyy{}\right)\,,
\end{equation}
and originates from compressibility terms of the form $\nabla\cdot(1/B^2) \vec
  B\times\nabla$. 
Note that $z$ takes values in the range $[-\pi:\pi]$ and that the outboard
mid-plane is located at $z=0$.
The perpendicular Laplacian is in the locally shifted metric
\cite{Scott:1997:2} written as 
\begin{equation}
  \nabla_{\perp}^{2} = \left(\ppxx{}+\ppyy{}\right)\,,
\end{equation}
and is due to $-\nabla\cdot\left(B^{-2} \vec B\times \vec B\times \nabla
\right)$, thus hiding magnetic shear in the shifting procedure.
The viscous/diffusive terms $\sim \mu_{\vor},\mu_n$ in Eqs.~(\ref{eq:eqvor})
and (\ref{eq:eqne}) are introduced to provide sub-grid dissipation of small 
scale dynamics.                    
\\
The parameters in the equations reflect the competition between parallel
and perpendicular dynamics, governed by the scale ratio
$\epss=(qR/L_\perp)^2$.  The electron parallel dynamics is controlled by
\begin{equation}
\label{eq:eqparms}
  \bhat={2 \mu_0 p_e\over B^2}\,\epss\,, \qquad\qquad
  \muhat={m_e\over M_i}\,\epss\,, \qquad\qquad
  C = 0.51 \frac{L_\perp}{\tau_e c_s}\muhat = \wh \nu \muhat\,,
\end{equation}
where $\tau_e$ is the electron collision time
and the factor  $0.51$ reflects the parallel
resistivity
\cite{Braginskii:1965}.  The competition between these three parameters,
representing magnetic induction, electron inertia, and
resistive relaxation, determines the response of $J$ to the static force imbalance in
Eq.~(\ref{eq:eqpsi}).
Due to the presence of $\ddpp$ in
Eq.~(\ref{eq:eqampere}) this adiabatic response has different character
in different parts of the spectrum.  The last physical parameter
is $\wcv$ in Eq.~(\ref{eq:eqcv}),
reflecting the effects of magnetic curvature (equivalently
magnetic gradient, in a toroidal model).  An important note is that
all magnetic induction $\ppt{\Psi}$ and flutter $\hat \beta \{\Psi, \cdot \}$ effects
enter through the finite beta $\beta = 2 \mu_0 p_e/B^2$
or $c_s^2/v_A^2$, where $v_A$ is the Alfv\'en velocity, and
$\bhat=\beta_e\epss$.
\\
The density equation is augmented by damping layers in the left and
right 5\% of the radial domain, regulating the poloidally averaged
density, e.g. the profile modification, back to zero. This feedback
control of the profile arranges for the average density profile to
stay close to the one characterized by the originally defined gradient.

\section{Energetics and evolution of flows \label{sec:energy}}

The equation determining the evolution of zonal flows is found from the
vorticity equation Eq.~(\ref{eq:eqvor}) by averaging over a flux surface as
\begin{equation} \label{zfeq}
\frac{\p V_0}{\p t} + \frac{\p}{\p x}\langle v_x v_y\rangle -
\wh{\beta}\,\frac{\p}{\p x}\langle B_x B_y \rangle + \omega_B\langle n\sin z \rangle = 
\mu_{\vor}\,\frac{\p^2V_0}{\p x^2} \, ,
\end{equation}
where $\langle \cdot \rangle = (1 / 2 \pi L_y) \int_{-\pi}^\pi
 dz \int_0^{L_y} dy \cdot$ denotes the flux surface average.   
The $E \times B$ velocity is given by ${\bf v_E} = (v_x,v_y,0) = (-\p_y \phi,
\p_x \phi, 0)$ and the electric field connected to the poloidal flow is
described by the potential  $\Phi_0(x) = \langle \phi \rangle$. 

Consequently $V_0(x) = \langle v_y \rangle = \p_x \Phi_0$ and $\langle \vor
\rangle = \p_x V_0$.
The first contribution to the flow evolution is the Reynolds stress, which
is the radial transport of poloidal momentum by radial velocity
fluctuations. It demands a correlation between the two components of the
fluctuating velocity, which may be due to a seed flow or a background
gradient, as is the case here.
The second contribution arises from magnetic flutter. It can be interpreted as
parallel current flowing radially along perturbed magnetic field lines. The
third term is the acceleration of the flow due to interaction with density
sidebands via the compressibility of the diamagnetic drift, associated with
geodesic acoustic modes. Finally, viscosity on the right hand side of
Eq.~\ref{zfeq} introduces a damping of the flow profile. 
\\
We are interested in the energetics of the flow evolution.
To find the evolution of the energy in the mean flow we multiply the
vorticity equation Eq.~(\ref{eq:eqvor}) by the flow velocity and
integrate over the whole volume.
We then obtain for the time evolution of the mean flow energy $U :=
(1/2)  \int d\vek{x}  V_0^2  $: 
\begin{equation}\label{eq:mean_energy}
  \frac{d U}{d t} ={\cal R} + {\cal M} + {\cal G} + {\cal V }\, ,
\end{equation}
with the quantities ${\cal R}, {\cal M} ,{\cal G},$ and $ {\cal V }$
defined as follows: From the convection we find
\begin{eqnarray}
{\cal R} = \int d\vek{x} \Phi_0 \langle {\bf v_E} \cdot \nabla_\perp \vor
    \rangle  
    & = & \int d\vek{x}  \langle v_x v_y \rangle \p_x V_0
    \label{eq:convective} . 
\end{eqnarray}
which is the Reynolds stress contribution to the flow drive. 
Correspondingly the parallel current and magnetic fluctuations lead to
\begin{eqnarray}
{\cal M} = \int d\vek{x} \Phi_0 \langle  \nabla_{\|} J  \rangle 
    & = & - \hat{\beta} \int d\vek{x} \langle {B}_x {B}_y
    \rangle \p_x \ V_0\; ,  \label{eq:emdrive} 
\end{eqnarray}
which is the Maxwell stress governing the energy exchange of the flow with
magnetic fluctuations, where ${B}_x = \p_y A_\parallel$ and ${B}_y = ´-\p_x
A_\parallel$.  
While the contribution of the normal curvature vanishes, the geodesic
curvature results in a term associated with the geodesic acoustic modes (GAMs):
\begin{equation}
{\cal G} = \int d\vek{x} \Phi_0 \langle \omega_B \sin s \p_x n \rangle 
     =   -\omega_B \int d\vek{x} \langle n V_0 \sin s \rangle  
     \label{eq:GAMS}. 
\end{equation}
The collisional damping finally is always  a sink:
\begin{eqnarray}
{\cal V } = -\mu_{\vor} \int d\vek{x} \Phi_0 \langle  \nabla^2_\perp \vor
  \rangle  
  & = &  -\mu_{\vor} \int d\vek{x} (\p_x V_0)^2  .\label{eq:kin}
\end{eqnarray}

Flow generation by Reynolds stresses is well known to result from an average
phase correlation between the velocity fluctuations in the drift plane spanned
by the $x$ and $y$ coordinate axes. The tendency of convective structures to
be tilted with a seed sheared flow makes the transfer term $\cal{R}$ generally
positive, draining energy from the fluctuating motions to the zonal
flows~\cite{Bian:Garcia:2003}.

It is worthwhile to note that in pure MHD turbulence there is an approximate
balance between Maxwell and Reynolds stress \cite{Kim:Hahm:Diamond:2001}. 
From a local linear analysis of modes, neglecting the toroidicity of the
equilibrium magnetic field, we obtain the following functional relationship
between the fluctuations in magnetic potential and electrostatic potential:
\begin{equation}
A_\| = \frac{(\omega_B k_y) / (k_\| k_\perp^2) + c}{
[\omega_B k_y c (\hat \beta - \hat \mu k_\perp^2)] / [ k_\|  k_\perp^2] + 1}
\phi \;,  
\end{equation}
with $c = \omega/k_\|$. The dispersion relation has several branches
(see Scott \cite{Scott:1997:1}).  In the limit of high $\hat \beta$ and
neglecting effects of curvature, the Alfv\'en branch of the dispersion
relation dominates and $c$ can be approximated by the Alfv\'en speed
$v_A = {\hat \beta}^{-1/2}$:
\begin{equation}
\label{eq:alfven}
A_\| =  \phi / \sqrt{\hat \beta}\, .
\end{equation}
As a consequence the Maxwell and Reynolds stress cancel in that
regime, which is expressing the fact that Alfv\'en waves do not
transport poloidal momentum.

From the plasma continuity equation~\eqref{eq:eqne} we find the evolution of
the density sidebands, 
\begin{equation} \label{dsbeq}
\frac{\p}{\p t}\langle n\sin z \rangle + \frac{\p}{\p x}
\langle \sin z\:n\,\frac{\p\phi}{\p y} \rangle +
\omega_B \langle \sin ^2 z\,\frac{\p n}{\p x} \rangle =
\omega_B \langle \sin ^2 z\,\frac{\p\phi}{\p x} \rangle -
\langle \sin z\,\frac{\p u}{\p z} \rangle .
\end{equation}
The contribution from the flow $V_0=\p\phi_0/\p x$ in the first term on the
right hand side of Eq.~\eqref{dsbeq}, describing the up-down asymmetric
plasma compression due to poloidal rotation, couples with the zonal
flow equation~\eqref{zfeq} and results in geodesic acoustic modes
(GAMs) at frequency $\omega_B/\sqrt{2}$
(Refs.~\cite{Winsor:Johnson:Dawson:1968, Hallatschek:Biskamp:2001, 
 Hassam:Drake:1993, Guzdar:Drake:McCarthy:Hassam:Liu:1993}). 
Other terms in Eq.~\eqref{dsbeq}, along with coupling to the ion flow
sidebands, may cause an acceleration of zonal flows in the presence of
poloidally asymmetric particle fluxes, known as Stringer-Winsor spin-up
\cite{Hassam:Drake:1993, Guzdar:Drake:McCarthy:Hassam:Liu:1993,
  Hallatschek:Biskamp:2001,Stringer:1969}. 
In this connection we also note that the energy transfer due to toroidal
geometry  into the energy of the fluctuating motions  
\begin{equation}
\label{Eq:kinEnergy}
{\cal K} = \frac{1}{2} \int d\vek{x} \; \tilde v ^2 \, ,
\end{equation}
is given by
\[
- \int d\vek{x}\,\wt{\phi}\mathcal{K}(n) = -
\omega_B\int d\vek{x} \left( \sin s\:n\,\frac{\p\phi}{\p x} +
\cos s\:n\,\frac{\p\phi}{\p y} \right) .
\]
This indeed indicates the tendency towards a ballooning structure
of the fluctuations, since this term drives velocity fluctuations when
the turbulent plasma transport is radially outwards from the torus
axis and poloidally towards the out-board mid-plane. This geodesic
transfer process was recently revisited in Ref.~\cite{Scott:2003},
where it was claimed that the GAM transfer is generally from the zonal
flows through the density side-bands to the turbulent fluctuations.

\section{Structure of electromagnetic fluctuations\label{sec:resglobal}}

To address the simultaneous action of the energetic transfer effects
we resort to three-dimensional numerical computations of the four-field
model eqs.~\eqref{eq:eqvor}- ~\eqref{eq:equi} on a grid of usually 
$64\times256\times32$ points, with dimensions $64\times256\times 2\pi$ in $x$,
$y$ and $z$, respectively. 
Some runs were repeated at higher resolution $128 \times 512\times 32$
to ensure convergence. The numerical scheme uses a symmetry, energy and
vorticity conserving discretisation of the bracket structure of the
nonlinearities \cite{Arakawa:1966} with the curvature terms cast into
bracket form as well. Time stepping is performed using an explicit third order
stiffly-stable scheme \cite{Karniadakis:Israeli:Orszag:1991}, with viscous
terms treated implicitly using operator splitting. For more details on the
numerical implementation see \cite{Naulin:2003}.

Nominal parameter values typical for tokamak edge plasmas are
$\wh{\epsilon}=18750$, $\wh{\mu}=5$, $\hat{s}=1$, $\omega_B=0.05$ and
$\mu_{\vor}=\mu_{n}=0.025$. 

\begin{figure}[h!]
\centering
\includegraphics[width=12cm]{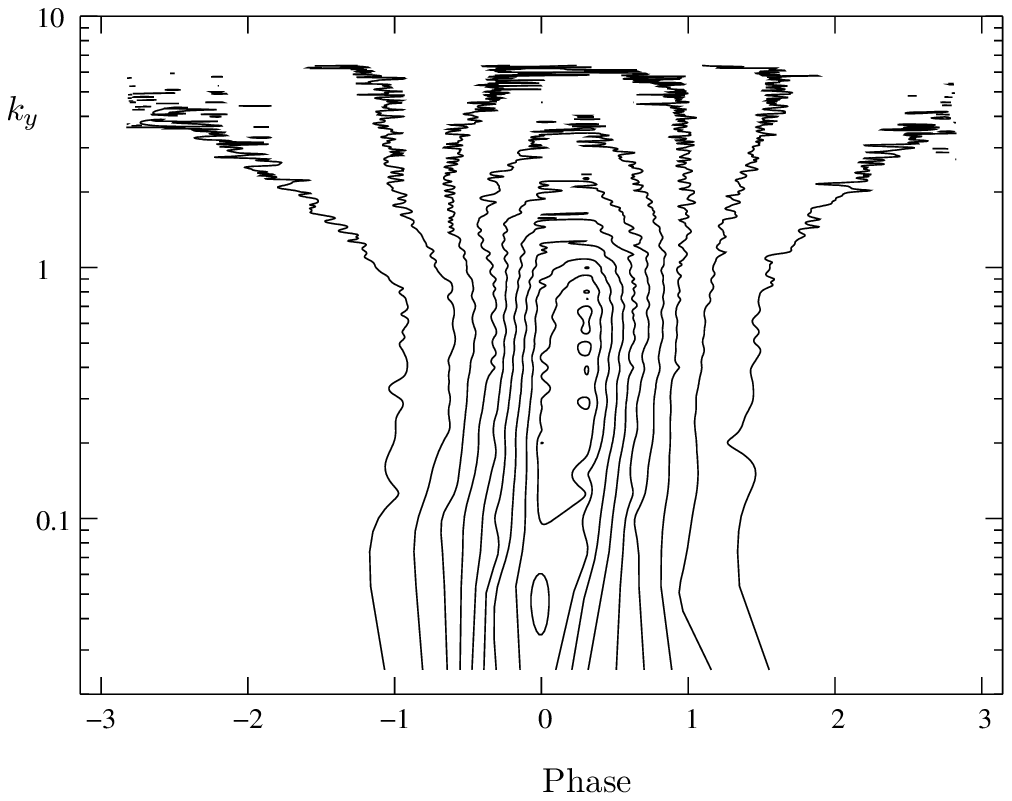}\\
\includegraphics[width=12cm]{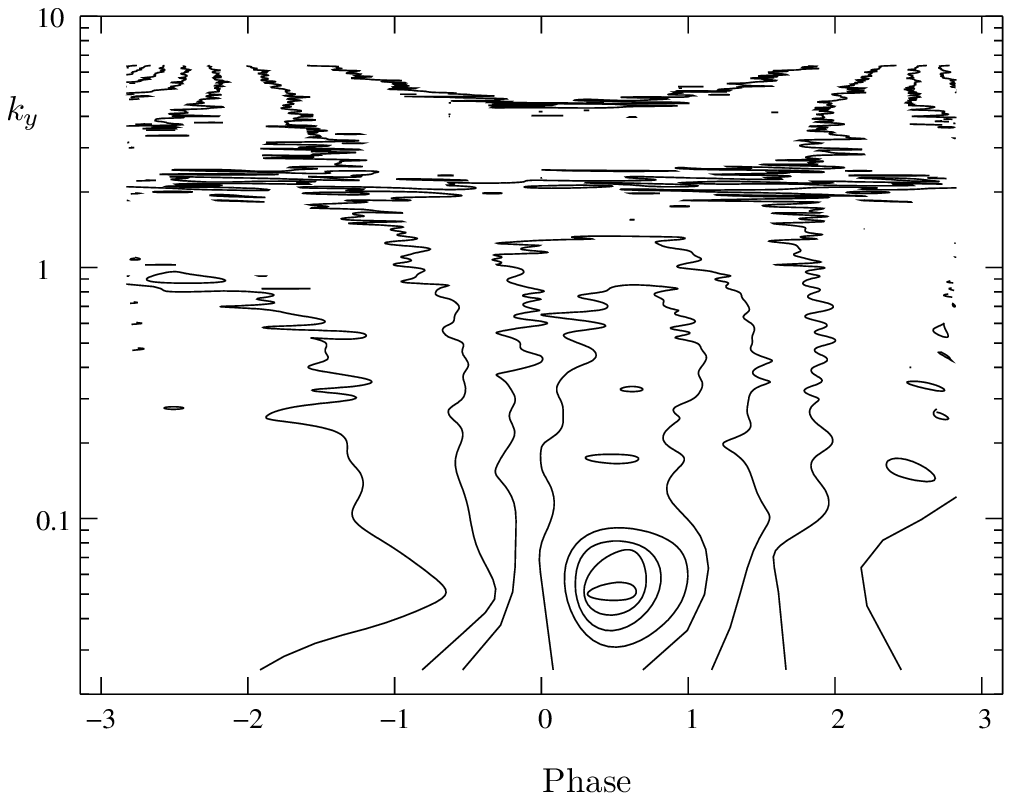}
\caption{Phase angle probability between density and potential
  fluctuations for the low  $\hat \beta = 0.1 $ (top) and high $\hat
  \beta = 30$ (bottom) case. 
\label{Fig:Phases}}
\end{figure}

For parameter scans we varied  $\wh{\beta}$ in a range between
$0.1$ -- $30$  and $\wh{\nu}$ from $0.5$ -- $7.5$.
The scaling with $\wh{\beta}$ is especially interesting, as 
the drift-Alfv{\'e}n system has the  property that the 
nature of the turbulence changes with the degree to which the
system is electromagnetic. This feature was demonstrated
numerically by Scott \cite{Scott:1997:1} and Naulin \cite{Naulin:2003} and also
experimentally by Lechte {\sl et al} \cite{Lechte:Niedner:Stroth:2002}:   
The transition manifests itself in a change of the phase relationship
between density and potential fluctuations, which varies for low values of
$k_y$ from a small phase angle in pure drift wave dynamics to $\pi/2$ in the
MHD drift-ballooning regime. 
This is exemplified in Figure \ref{Fig:Phases}, which shows the phase
probability distribution function as function of poloidal wavenumber
for the cases $\hat \beta = 0.1$, $\wh \nu = 2.295 $ and $\hat \beta =
30$, $\wh \nu =0.5$.
While in the low $\hat \beta$ case, the phase angle is always small,
for the large $ \hat \beta$ cases we observe a much broader phase
relationship and a generally larger phase angle. The regime of
dominating MHD ballooning instability is first reached at for the
edge very  high values of $\hat \beta >  30$ \cite{Naulin:2003}.

\begin{figure}[h!]
\centering
\includegraphics[width=12cm]{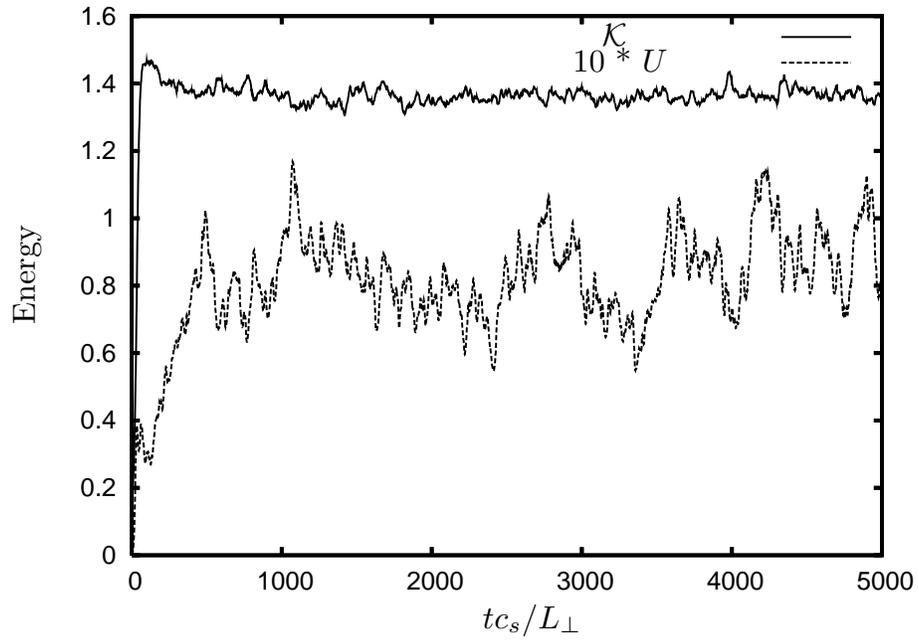}
\caption{Kinetic energy $\cal K$ and energy in  zonal-flow component
  $U$  over time for the low $\hat \beta = 0.1$ case.
\label{Fig:EnergyTime}}
\end{figure}

A time-trace of the kinetic energy, $\cal K$, of the fluctuating motions
and the zonal flow energy $U= \int d \vek{x} (1/2) V_0^2$ is presented in
Fig.~\ref{Fig:EnergyTime}. It is seen that while $\cal K$
saturates after about 100 time units, the saturation of the flow takes
place much later. Thus, all computations were run to times $t= 5000$,
with time averages taken in the interval from time $t=1000$ to the end of
the simulation, to ensure a statistical steady state of fluctuating quantities.
Moreover,  from Figure \ref{Fig:EnergyTime} it is observed that the energy in
the zonal flows is only a fraction of the total kinetic energy. 
This underlines the fact that no strong self-organized H-mode like transport
barriers are formed in this system. However, zonal flows do form and influence
the profile of transport as well as the density profile.  
Due to the change in turbulence character connected to $\hat \beta$ we
will now proceed and present in more detail two runs, the low beta $\hat \beta
=0.1$, $\wh \nu = 2.295$ and a high beta $\hat \beta = 30 $, $\wh \nu
= 0.5$  case. 

\begin{figure}[h!]
\centering
\includegraphics[width=12cm]{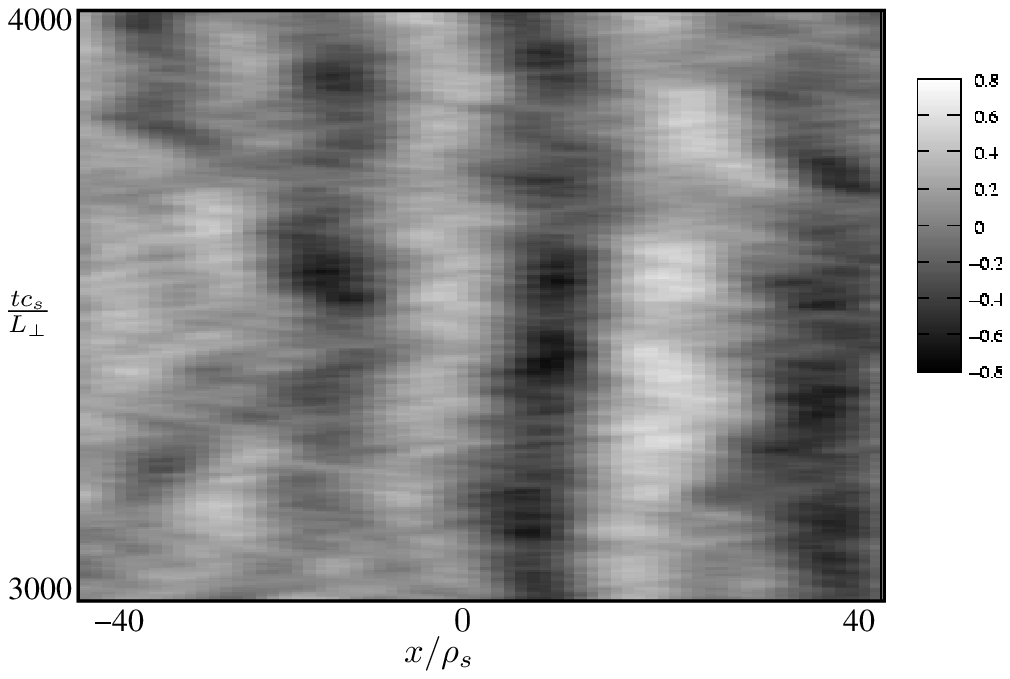}\\
\includegraphics[width=12cm]{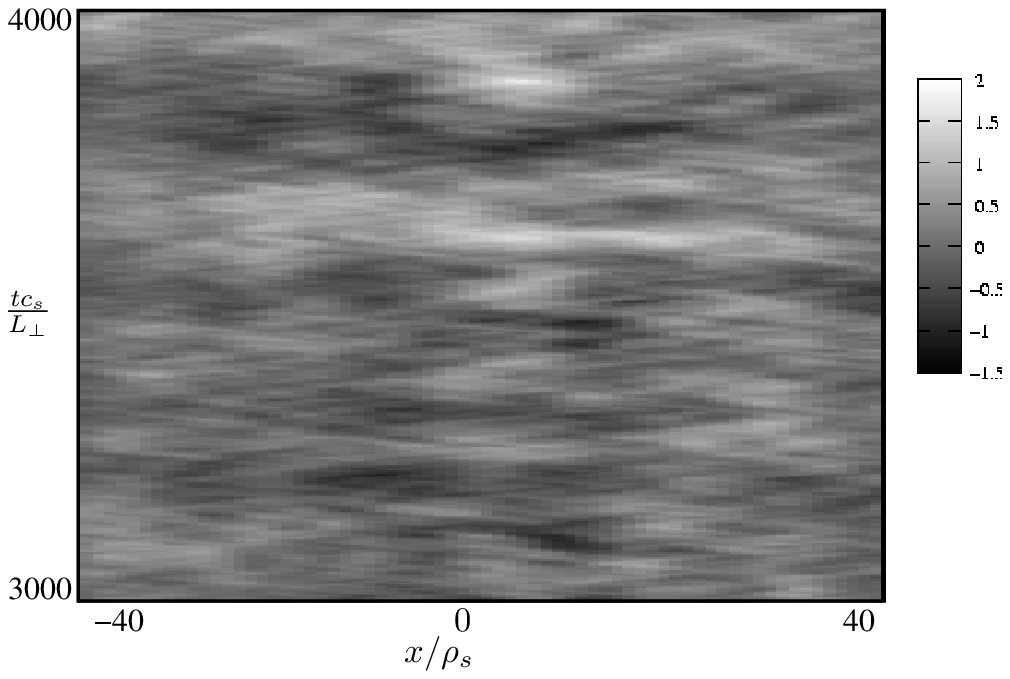}
\caption{Space-time evolution of zonal flow $V_0(x,t)$ (top) and zonal density
  $\langle n \rangle (x,t)  $ (bottom) for $\hat \beta = 0.1.$
\label{Fig:FlowTime_b0.1}}
\end{figure}

Figure \ref{Fig:FlowTime_b0.1} shows a gray-scale plot of the zonal
flow profile $V_0(x,t)$ and the zonal density $\langle n \rangle (x,t)$
in time, where in both cases we omit the damping layers in the plot. 
It is clearly seen that the zonal flows are radially localized and while
exhibiting some fluctuation features, the flow profile is rather persistent in
time. The zonal density shows some imprint of the zonal flow in terms of
slightly elevated density levels in the vicinity of high flow shear, but
fluctuations in the zonal density  are more pronounced. It is worthwhile to
remark that at about $t \approx 3750$ a weakening of zonal flows is clearly
noticeable and prompts a transient radially propagating feature in the zonal
density. This provides us a visualisation of the interplay between flows,
transport and the density profile.  

\begin{figure}[h!]
\centering
\includegraphics[width=12cm]{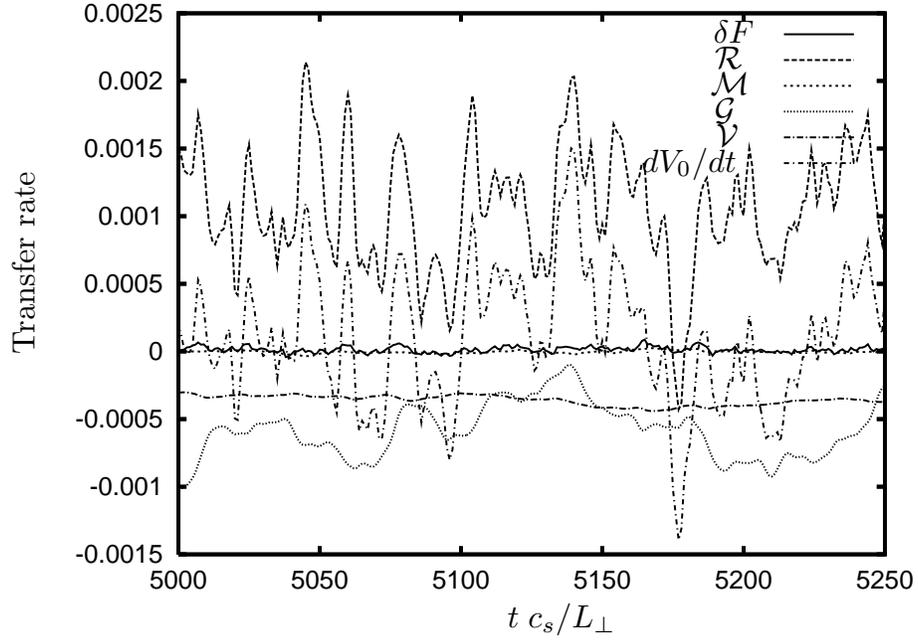}
\caption{ Energy exchange terms, flow change rate, and numerical error for
  $\hat \beta = 0.1$. Reynolds stress is the key drive and GAMs are
  acting as a sink for the flow energy.
\label{Fig:FlowGenBeta1}}
\end{figure}

\begin{figure}[h!]
\centering
\includegraphics[width=12cm]{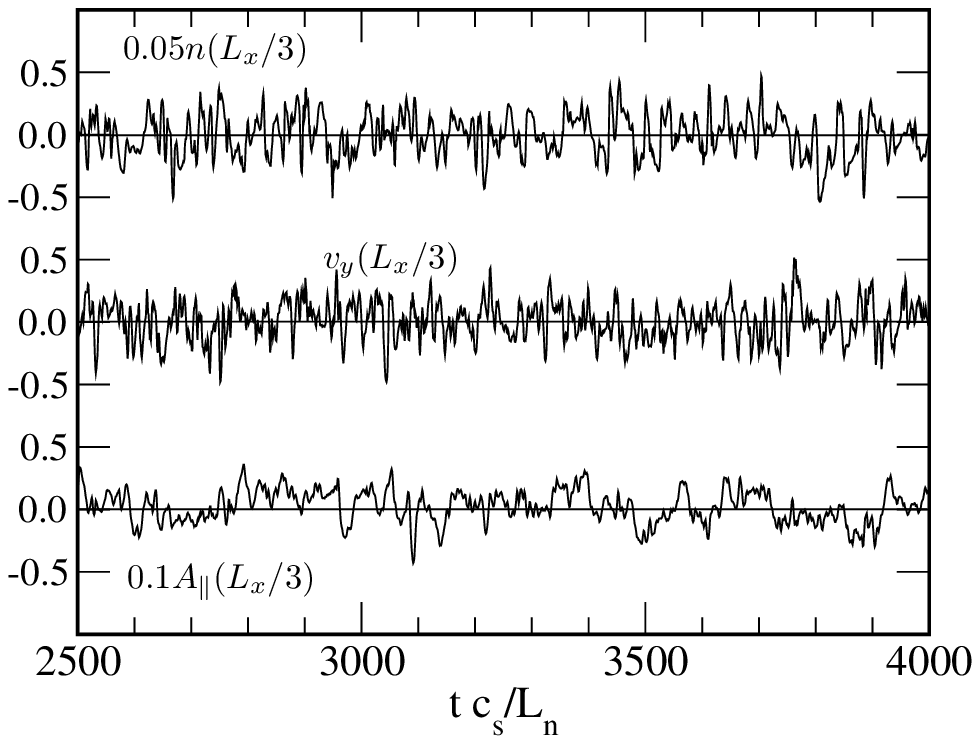}\\
\includegraphics[width=12cm]{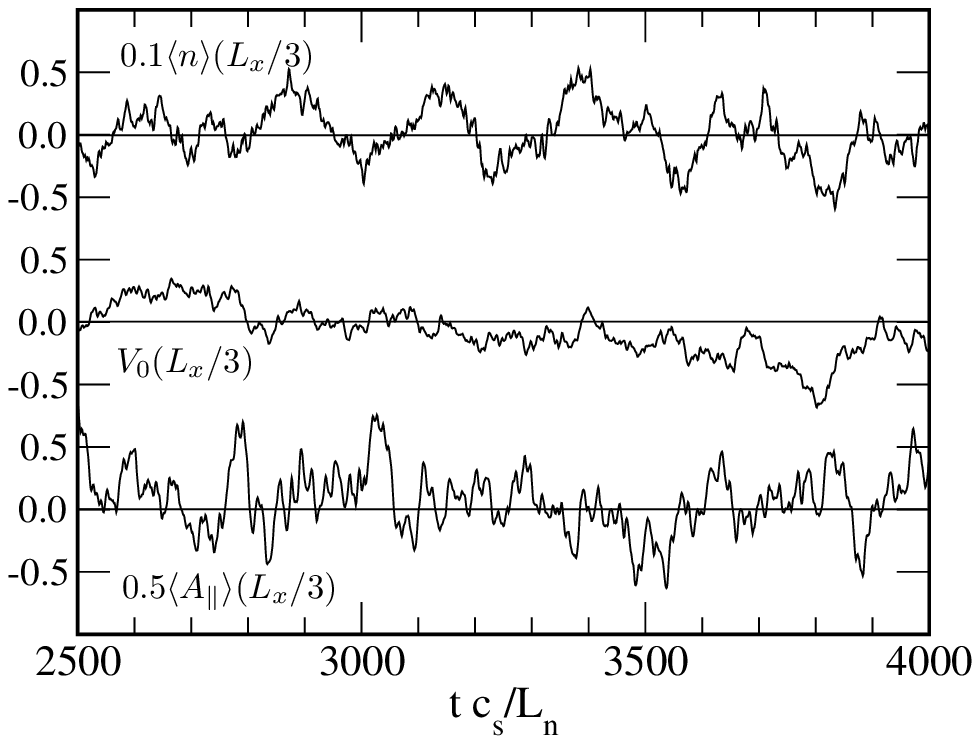}
\caption{  Fluctuating quantities (top) and fluxsurface
  averaged quantities (bottom) for $\hat \beta = 0.1$, measured at $x = L_x/3$  and on
  the outboard midplane. 
\label{Fig:slow:fast_b0.1}}
\end{figure}

\begin{figure}[h!]
\centering
\includegraphics[width=12cm]{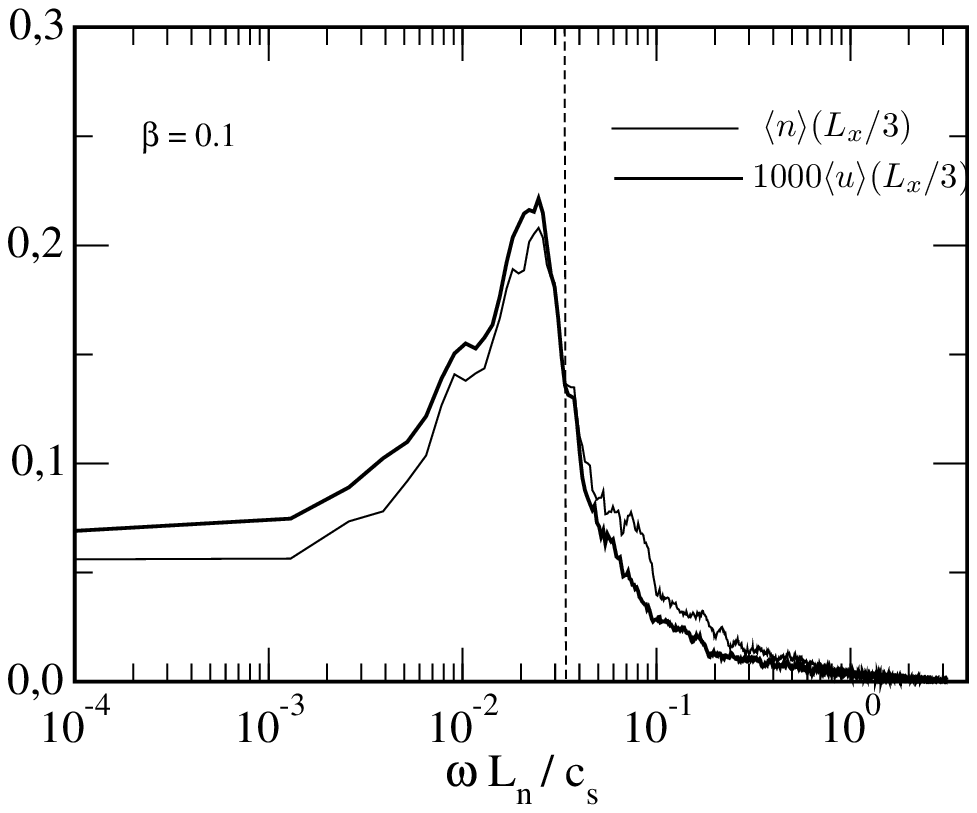}\\
\includegraphics[width=12cm]{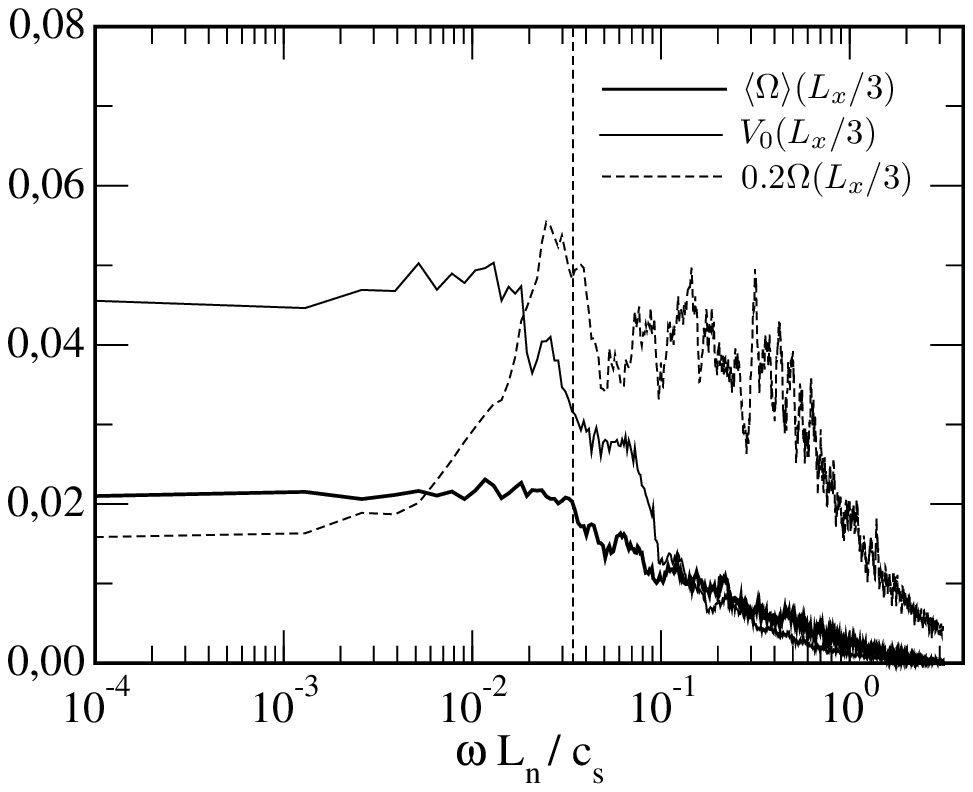}
\caption{  Frequency spectra of quantities
  associated with GAM oscillation (top) and flows (bottom) for $\hat
  \beta  = 0.1$. The
  vertical line indicates the ideal GAM frequency $\omega_{GAM}$.  
\label{Fig:Spectrabetya0.1}}
\end{figure}

In Fig.~\ref{Fig:FlowGenBeta1} we present time traces of the energy
transfer terms into the zonal flow and the rate of change of the zonal flow,
together with the numerical error obtained by comparing the sum of the energy
transfer terms with the computed actual change rate of the flow:
\begin{equation}
\label{Eq:Error}
 \delta F(t) =   \frac{dV_0}{dt}_{Num.} - \left( {\cal R} + {\cal M} + {\cal G} +
 {\cal V } \right)\;. 
\end{equation}
Here $ {dV_0}/{dt}_{Num.}$ is evaluated to second order in time \cite{Naulin:Nielsen:2003}.
The error $\delta F(t)$ is of the same order, and
for $\hat \beta =0.1$  the Maxwell stress energy
transfer term $ \cal M$, which is negligible for this low value of
$\hat \beta$, 
when compared to each other energy transfer term, is of the same size.
The statistical nature of the fluctuating flow drive terms is observed, as the
balance between the transfer terms is only reached on a long time average,
whereas on an instantaneous view the transfer terms can deviate significantly
from their means. 
Here viscous damping $ \cal V$ and GAMs $\cal G$ serve as sinks for the flow
energy, which is solely driven by the Reynolds stress. 
While both damping terms vary on a rather slow time scale, the Reynolds stress
and with it the resulting rate of the flow change vary on the faster time
scale of the turbulence. Figure \ref{Fig:slow:fast_b0.1} shows correspondingly
time traces of selected fluctuating quantities obtained at a single point and
of flux surface averaged quantities at the same radial position.
While the fluctuating quantities are all varying on the fast scale, the flux
surface averaged ones vary significantly slower. Zonal density and zonal
magnetic potential show, however, variations on a time scale of about 250 and
80 time units, respectively. The zonal flow $V_0$ shows some fast scale
jitter, but varies only slowly over the shown 1500 time units. 
Thus, to investigate that time behavior in more detail we  present in
Fig.~\ref{Fig:Spectrabetya0.1} the frequency spectra of flux surface averaged
quantities connected to GAMs, as the density and the parallel ion velocity
together with quantities related to zonal flow dynamics. 
The zonal density $\langle n \rangle$ and the flux-surface averaged parallel
ion velocity $\langle u \rangle$ show both a pronounced peak at a low
frequency of about $\omega  \approx 0.025$. This peak is clearly
associated to the ideal geodesic oscillation around $\omega_{GAM} \approx 
\omega_B/\sqrt{2} = 0.035$.  It is worthwhile to note that the GAM
frequency arises by combining Eq.~(\ref{zfeq}) with Eq.~(\ref{dsbeq}), and the
ideal GAM frequency arises from the relation
\begin{equation}
\omega_B \langle \sin ^2 z\,\frac{\p\phi}{\p x} \rangle 
= {1 \over 2} \omega_B \langle [1-\cos(2z)] \,\frac{\p\phi}{\p x}\rangle 
\approx {1 \over 2 } \omega_B V_0
\end{equation}
if the flux surface average  of the term $(\cos(2z) \,\partial_x \phi)$
disappears exactly, as it would be expected for a fluctuations 
$v_y$ being homogeneous along the parallel coordinate $z$. In toroidal
geometry the  $\phi(z)$ and thus $v_y(z) = \partial_x
\phi$, however,  show in general a distinct ballooning feature,
resulting in  higher amplitudes
around position $z=0$ than for $z=\pm \pi$.  
We thus expect the GAM peaks in the spectra to be shifted from the ideal
$\omega_{GAM} $, with the direction of the shift depending on the preferential
direction of local flows $v_y$, and the width of the frequency shift
depending on the ballooning properties of the velocity fluctuations
$v_y(z)$. For our present parameters and a probe location  one
third  into the
radial $x$ domain we experience a downshift by an additional factor of
approximately $\sqrt{1/2}$. 

For the zonal flow $V_0$ and the zonal vorticity, we observe that the zero
frequency mode dominates the poloidal flow spectrum. At the frequency of the
zonal density feature we observe even a small  dip in the flow frequency
spectrum, this supporting the observation that GAMs are a sink for the flow
energy in that parameter regime. 

\begin{figure}[h!]
\centering
\includegraphics[width=12cm]{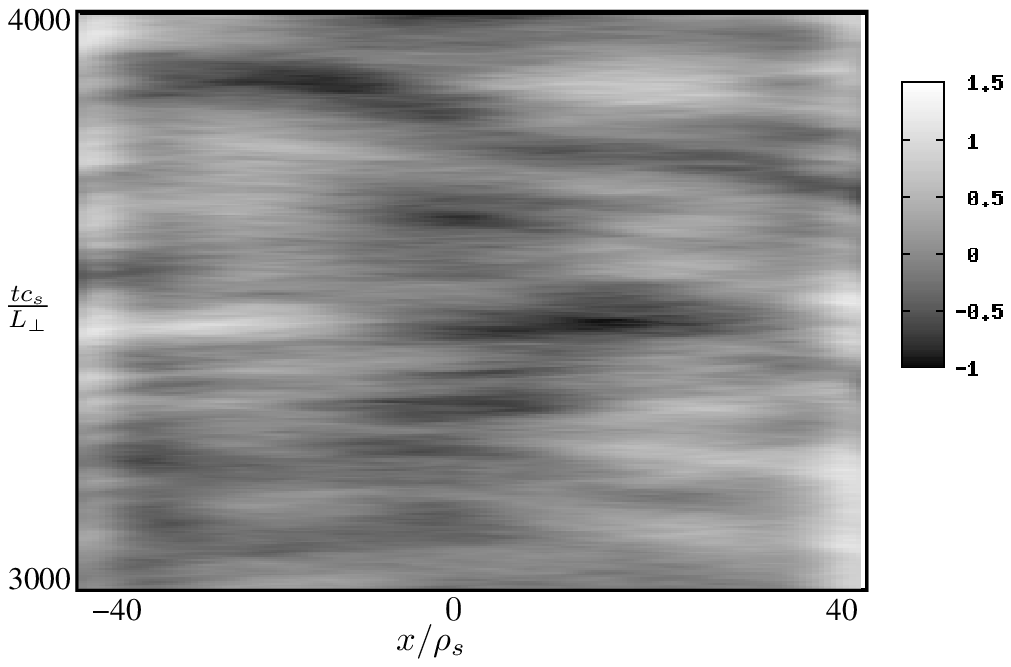}\\
\includegraphics[width=12cm]{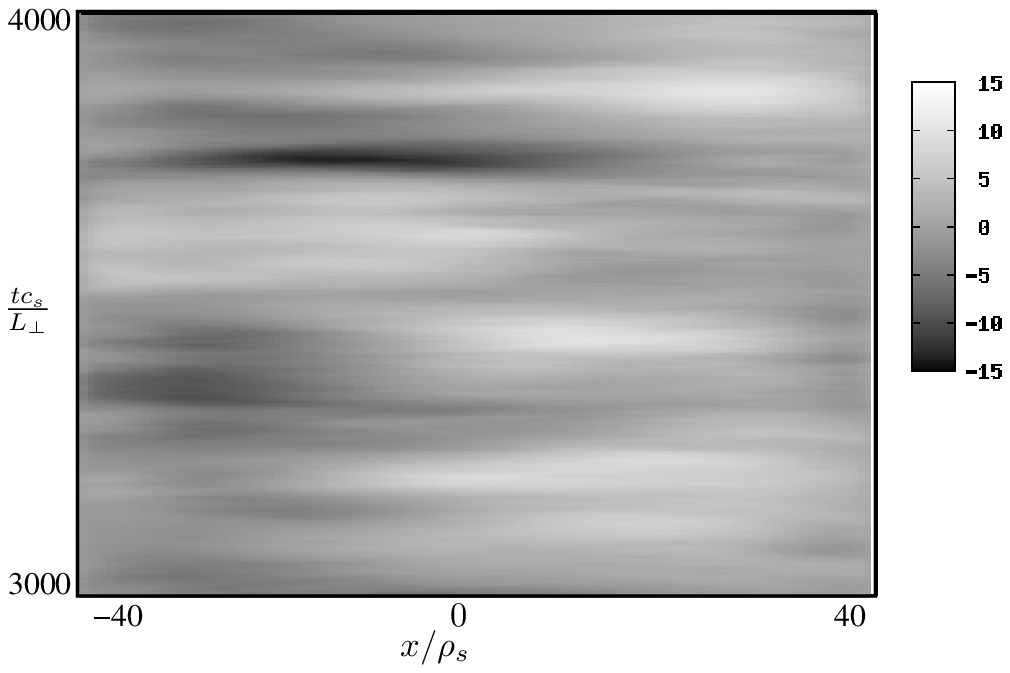}
\caption{Space-time evolution of zonal flow $V_0(x,t)$ (top) and zonal density
  $\langle n \rangle (x,t)  $ (bottom) for $\hat \beta = 30.$
\label{Fig:FlowTime_b30}}
\end{figure}

For the high $\hat \beta = 30$ case the situation looks differently
as is clearly seen from Fig.~\ref{Fig:FlowTime_b30}. The zonal flow
profile is now broader and reveals much less persistence than in the low
beta case (compare Fig.~\ref{Fig:FlowTime_b0.1}). Correspondingly the
zonal density shows also a less pronounced radial structure and the
characteristic time of the fluctuations seems to be of similar size for both
zonal flow and zonal density profile. 

\begin{figure}[h!]
\centering
\includegraphics[width=12cm]{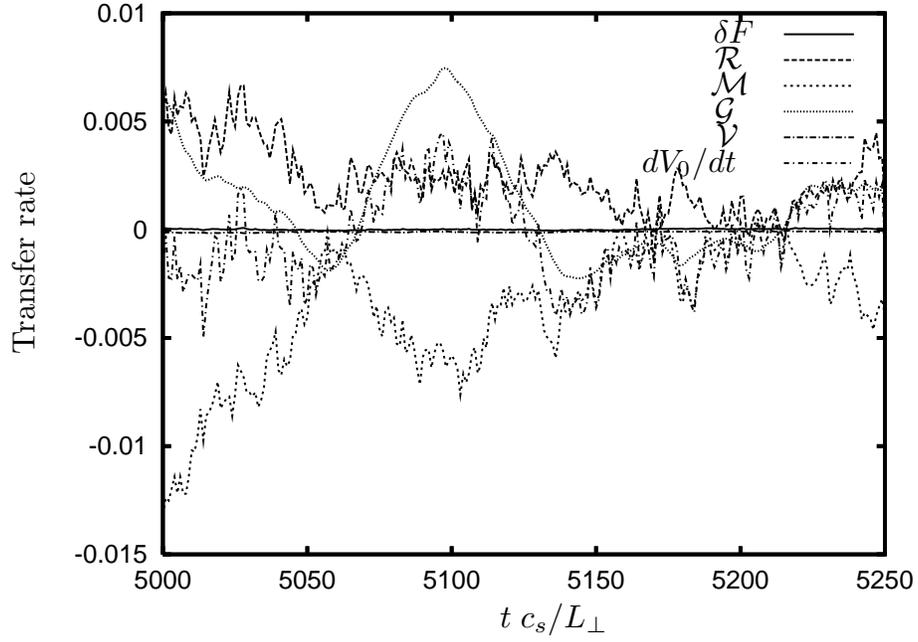}
\caption{ Energy exchange terms, flow change rate, and numerical error for
  $\hat \beta = 30$, showing the dominating influence of the GAMs for
  flow drive in that regime. 
\label{Fig:FlowGenBeta30}}
\end{figure}

The energy transfer rates $\mathcal{R}$, $\mathcal{M}$, and $\mathcal{G}$ ,
shown in  Fig.~\ref{Fig:FlowGenBeta30} reflect this change in behavior. 
We first observe that the Maxwell stress is now of finite size and a
significant sink for the flow energy. It is very well correlated to the
Reynolds stress in time, which still acts as a flow drive. 
For an ideal high beta MHD case in linear geometry without magnetic field
curvature the balance between $\mathcal{M}$ and $\mathcal{R}$ is known to be
exact with no preferred flow direction in the dynamics, as seen in
Eq.~(\ref{eq:alfven}). 
Here the resulting energy transfer from Maxwell and Reynolds stress is close
to zero and the resulting change in the effective zonal flow drive is
dominated by the GAMs $\mathcal{G}$, which now acts as a driving
term. Consequently the resulting rate of change for the flow evolves mainly
along with the change in the GAM drive and shows only a minor additional
variation on the timescale of the Reynolds and Maxwell stresses.

\begin{figure}[h!]
\centering
\includegraphics[width=12cm]{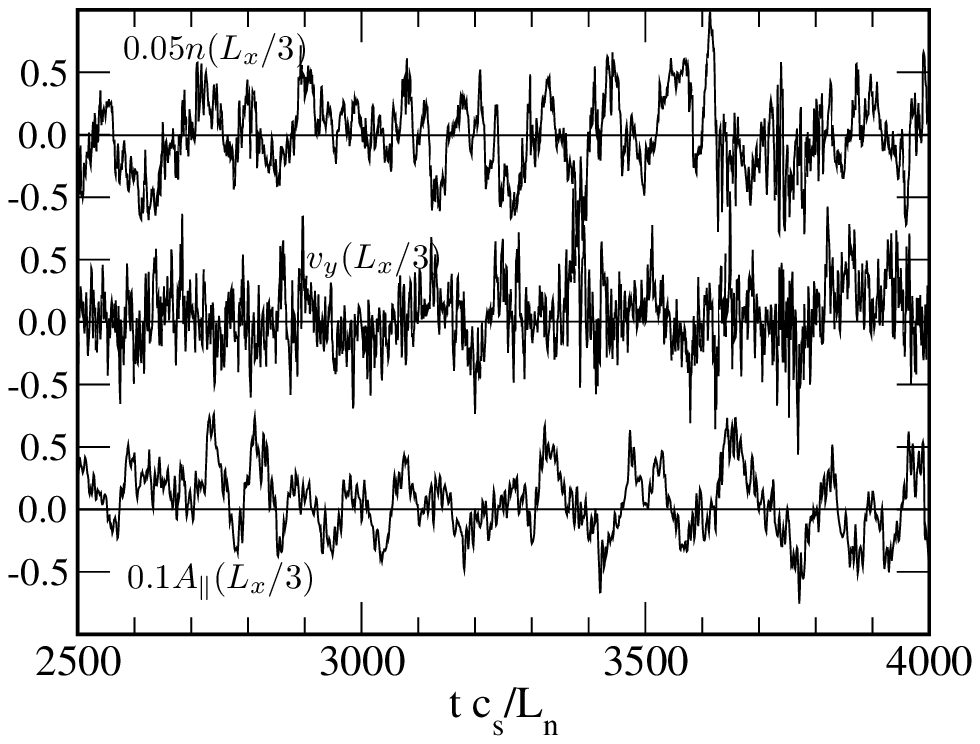}\\
\includegraphics[width=12cm]{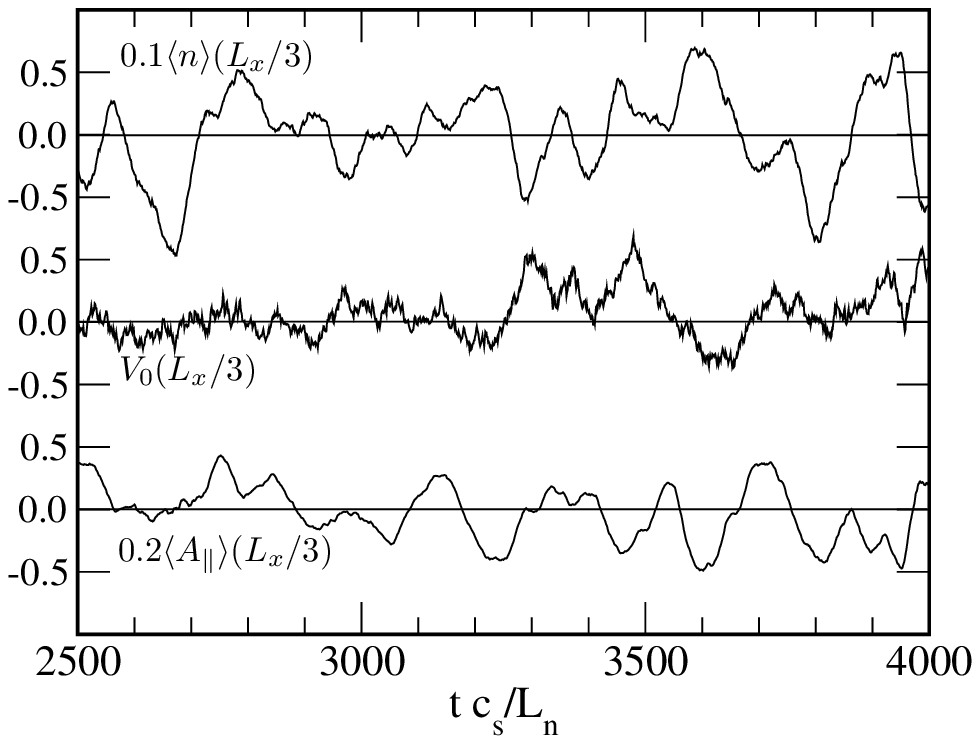}
\caption{ Fluctuating quantities and fluxsurface
  averaged quantities for $\hat \beta = 30$, measured at $x = L_x/3$ and on
  the outboard midplane. 
\label{Fig:slow:fast_b30}}
\end{figure}

\begin{figure}[h!]
\centering
\includegraphics[width=12cm]{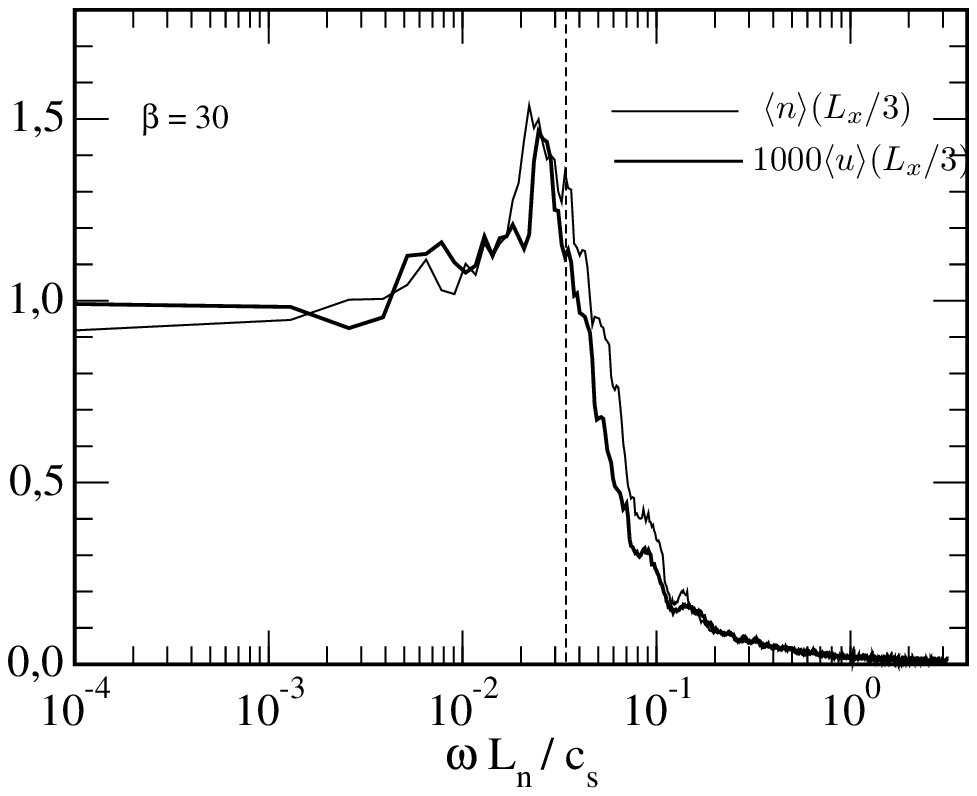}\\
\includegraphics[width=12cm]{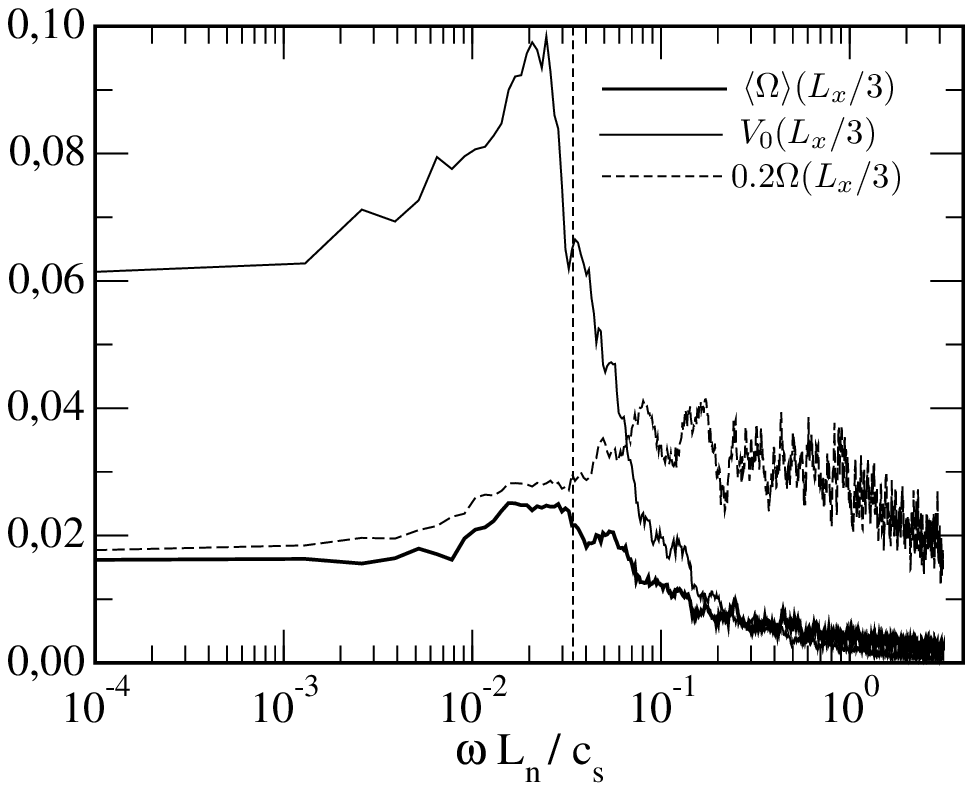}
\caption{  Frequency spectra of quantities
  associated with GAM oscillation (top) and flows (bottom) for $\hat
  \beta = 30$. The
  vertical line indicates the ideal GAM frequency $\omega_{GAM}$.  
\label{Fig:Spectrabetya30}}

\end{figure}

The time traces of fluctuating and averaged quantities shown in
Fig.~\ref{Fig:slow:fast_b30}, reflect that behavior. We first note 
that the fluctuating quantities now show a somewhat more pronounced
slower frequency, which reflects the frequency observed in the time
evolution of the zonal quantities. All fluctuations are larger by about a
factor three compared to the low beta $\hat \beta = 0.1$ case. The zonal
density is up by a factor two to threee  and a  slow oscillation is clearly
observed in both the zonal density as the zonal magnetic potential. 
This slow frequency now is also found in the zonal flow time trace. 
These features get more obvious in the frequency spectra depicted in
Fig.~\ref{Fig:Spectrabetya30}. 
A pronounced low frequency behavior is now seen also in the flow related
quantities, namely in $V_0$, which here exceeds the zero frequency component
by about a factor two.  
In these situations the flow is not stationary (zero frequency) but is,
compared to the turbulence, a slowly varying structure. The slow
frequency of the flow is close to the ideal GAMs frequency
$\omega_{GAM}$, revealing the flow drive by this process.
These results are generally in agreement with experimental observations, that
show a modulation of the zonal flows at the frequency of the GAM 
oscillation \cite{McKee:etal:2003, Conway:2004}.

\section{Plasma beta and collisionality scalings}

\begin{figure}[h!]
\centering
\includegraphics[width=12cm]{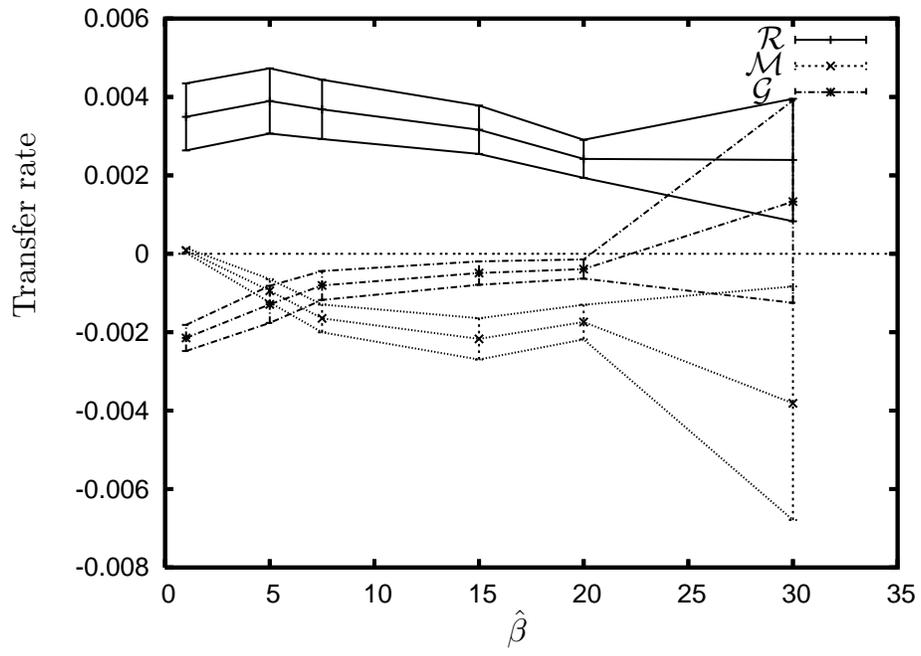}
\caption{Dependence of flow energy transfer terms  terms on $\hat
  \beta$ for $\wh \nu = 0.5$, with standard deviation.
\label{Fig:scan_beta}}
\end{figure}

Here we present results concerned with the scaling of the different
transfer terms with collisionality and plasma beta.
In Fig.~\ref{Fig:scan_beta}  we show the three main transfer terms
as a function of $\hat \beta$ for a low collisionality of $\wh \nu =0.5$. 
For increasing $\hat \beta$ the Reynolds stress drive $\cal R$ gets slightly
weaker, but is in all cases a drive. This indicates that the described flows
do not decay through a Kelvin-Helmholtz like instability mechanism, which
would make the Reynolds stress a sink term. The Maxwell stress $\cal M$ starts
close to zero and is always a sink term. It grows as expected in significance
with increasing $\beta$, and for large $\beta$ is the dominating sink for the flow
energy.
The GAM transfer, $\cal G$, starts out as a sink for the flow energy
at low beta, but with
increasing beta it looses its importance as a sink. 
Finally, $\cal G$  becomes positive for the high beta $\hat \beta = 30$ case, e.~g.~ the GAM acts as a flow
drive. 
  
\begin{figure}[h!]
\centering
\includegraphics[width=12cm]{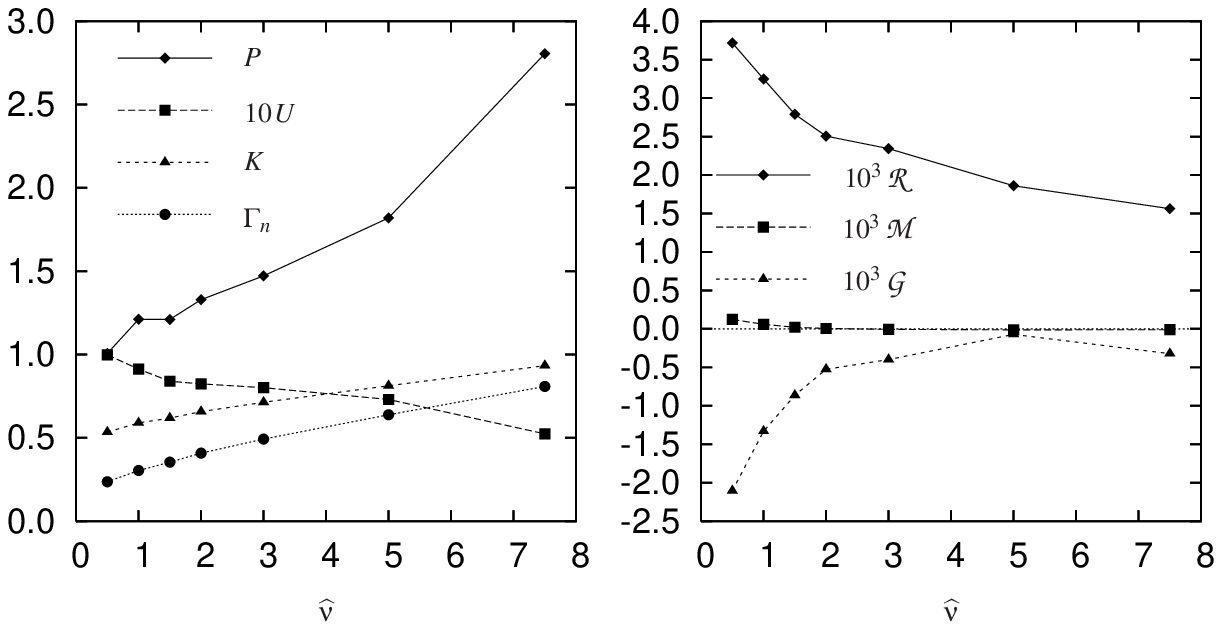}
\includegraphics[width=12cm]{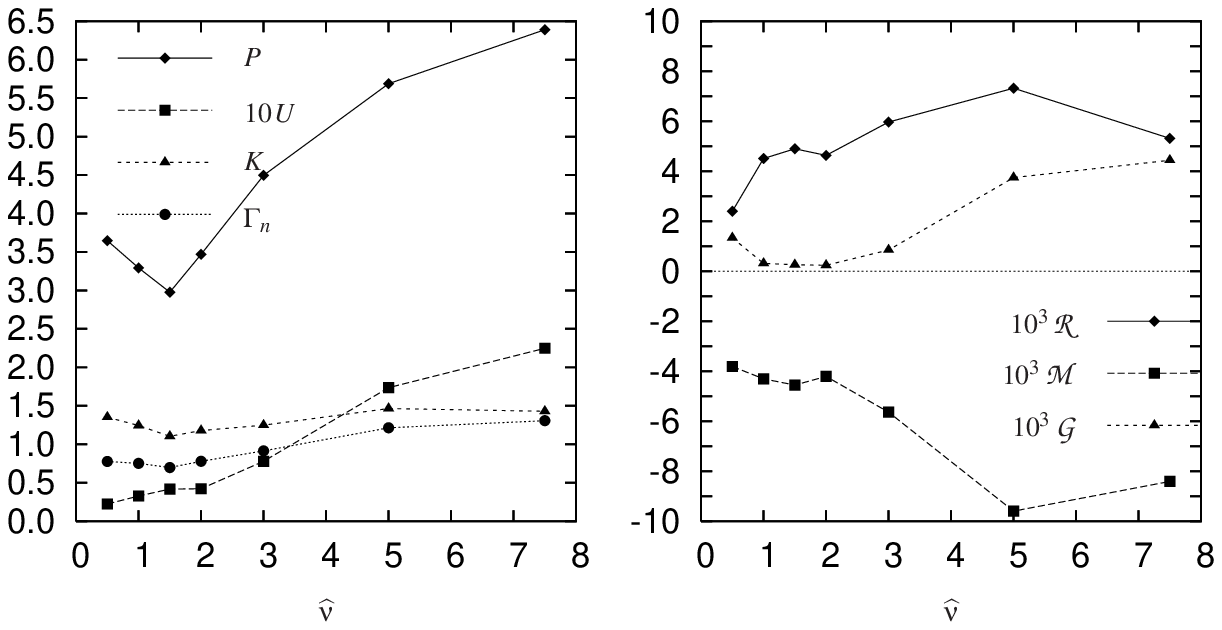}
\caption{Scan over collisionality $\wh \nu$ for low $\hat \beta = 1.0$ (top) and high 
$\hat \beta = 30$ (bottom). The left side shows energy in the density
fluctuations $P$, kinetic energy $K$, flow energy $U$ and particle
flux $\Gamma_n$. The right side depicts energy transfer terms.
\label{Fig:scan_nu_lb}}
\end{figure}

We then look at the scaling of flows and energy transfers with collisionality
in the two cases of low and high beta. The results are presented in Fig.~\ref{Fig:scan_nu_lb}. 
We find that the saturation level of zonal flow energy decreases with rising
$\wh{\nu}$ for low $\wh{\beta}$ and is in general by an order of magnitude
smaller than the fluctuating kinetic energy $\mathcal{K}= \int d {\bf x} (1/2)
(\nabla_\perp \tilde \phi)^2$. The system is mainly governed by vortex
dynamics \cite{Naulin:Garcia:Nielsen:Rasmussen:PLA:2004}, where the zonal flow contribution plays an important part for
self-regulation of the fluctuation amplitudes, but is not dominating the
energetics to an extent that it would completely suppress the turbulence. 
Further, we observe that the fluctuation energies $\mathcal{K}$ and
$\mathcal{P}=\int d {\bf x} (1/2) \tilde n^2$ and the turbulent particle transport
$\Gamma_n=\int d {\bf x} v_x  n$ both increase with parallel resistivity $\wh{\nu}$. This stronger
turbulence level is due to the increasingly non-adiabatic nature of the
electron response due to collisions, which increases the nonlinear drive of
drift modes.  
The energy transfer terms behave as follows: The Reynolds stress
decreases with increasing collisionality, accordingly with the
decaying flow energy. The Maxwell stress energy transfer $\cal{M}$ is
always negligible at low beta and thus the flow energy is dissipated
through the viscous terms $\cal{V}$  and the geodesic channel. 

Energies $\mathcal{K}$, $\mathcal{P}$ and the transport $\Gamma_n$  approximately
double when changing to $\wh{\beta}=30$ and taking the system from the drift
into the ballooning regime. 
The magnetic flutter effect is then an important cause for the non-adiabatic
response on the electrons. The influence of $\wh{\nu}$ is accordingly much
smaller and results in less variation  for this high $\wh{\beta}$ case. 

We observe that the Reynolds stress is always a drive
($\mathcal{R}>0$) for the flow. At high beta the Maxwell stress is important and the balance between Maxwell
and Reynolds stress becomes obvious as $|\mathcal{M}|\approx \mathcal{R}$ is
reached and exceeded. Finally the electromagnetic flutter effect is taking out more energy
from flows than is injected by Reynolds stress spin-up. The flow 
sustained in this high beta regime is now maintained by geodesic transfer
{\sl into} the flow. The driving effect by GAM oscillations on the flow is
more pronounced for higher $\wh{\nu}$ and higher levels of transport: 
GAM transfer is closely linked to the energy of density fluctuations in the
$m=\pm 1$ sideband (and thus to energy in all other scales that couple by
three-wave interaction to this sideband), which directly scales the transfer
term $\mathcal{G}$. For higher resistivity the relative importance of
$\mathcal{G}$ is thus enhanced in the same amount as $\tilde n$ 
increases both due to a more resistive as well as more
electromagnetic electron response.  
The strong magnetic flutter $\tilde B_\perp$ in high beta turbulence can cause
significant chaotic deviations of field lines radially and poloidally from the
flux coordinates that were defined for an unperturbed magnetic field. The
parallel coupling is thus  able to connect regions of neighboring radial
domains where the amount of radial overlap is rising with $\wh{\beta}$. Zonal
structures may be efficiently destroyed for a strong flutter effect, and
the aligning character of Reynolds stress on zonal flows is
counteracted, thus also the radial structure of the zonal flows is
less pronounced in the high beta case, see Fig.~\ref{Fig:FlowTime_b30}.
The drive of vortices on the drift wave scales is of course still maintained
by the free energy in the background density gradient, and the energy on
drift scale density fluctuations is even increased by the destabilising
magnetic flutter effect on the non-adiabatic parallel dynamics. The cascade in
density structures is generally a three-wave interaction that is on
statistical grounds essentially down to smaller scales, but is by more
infrequent events also able to feed scales in the $m=\pm1$ geodesic
sidebands and the $m=0$ zonal mode. The geodesic transfer pathway is thus open
in both directions: a strong drive of the GAM mode by zonal flows
for low beta on the average drains  energy out of the flow to smaller
scales.
For high beta the GAM energy is, however, converted to a certain extend into
$\tilde \phi$ oscillations that supply the zonal flows.

\section{Conclusion}
We have performed a detailed investigation of the zonal flow drive in
drift-Alfvén turbulence for parameters relevant to the edge region of
hot plasmas in toroidal devices. We have identified three main
mechanisms for the interaction of the zonal flows with the turbulent
fluctuations; namely the electrostatic Reynolds stress,  the
electromagnetic Maxwell stress, and the geodesic acoustic mode, GAM,
coupling. The main results are summarized as follows: For low beta
cases the Maxwell stress is negligible and the Reynolds stress is the
only driving term of the flow, whereas the GAM coupling provides a sink
for the flow in addition to the viscousity. For the case of high-beta
plasmas, however, the Maxwell stress becomes significant. It acts as a
sink for all the cases we have investigated, and it efficiently cancels the
driving effect of the Reynolds stress.  In this parameter regime the
flow is mainly sustained by the GAM coupling, that now acts as a drive
opposing the viscous damping.

We should therefore emphasize that from an experimental point of view,
measuring Reynolds stress exclusively as
an indication for flow generation is in general not sufficient. 
The electromagnetic Maxwell stress is important already at a moderate edge
beta parameter, and  will be even more important for ITER like plasmas with
higher edge $\beta$ at reduced collisionality. This clearly opens a demand for
additional measurements of the Maxwell stress. 

There is a clear trend in the computational results that assign more
importance to the GAM oscillation at a high level of transport. 
The GAMs present a driving mechanism for the flows if the transport is
sufficiently inhomogeneous: with an increasing ballooning character of the
turbulence the GAMs are further excited and can ultimately drive flows. 

The frequency spectra of the zonal flow clearly show a dip or a peak in the
$\omega_{GAM}$ frequency range, depending on the sink or drive role of the GAMs
for the flow evolution.  
Measurements of the frequency spectrum of the zonal flow should thus be able
to distinguish between these two scenarios and provide further insight into
the importance of GAMs for the flow, and finally for H-mode formation. 

Finally we note that our numerical results are for the high beta case
partially in disagreement with recent results by B. Scott \cite{Scott:2003}
regarding the specific role of the Maxwell stress \cite{Scott:NJP:2005}.

\acknowledgments
This work was supported by the Danish Center for Scientific Computing (DCSC),
grants CPU-1101-08 and CPU-1002-17. 

\newpage

\bibliographystyle{prsty}
\bibliography{}

\newpage
 
\section*{Figures}
 
\clearpage
\newpage

\clearpage
\newpage

\clearpage
\newpage

\clearpage
\newpage

\clearpage
\newpage

\clearpage
\newpage

\clearpage
\newpage

\clearpage
\newpage
  
\clearpage
\newpage

\clearpage
\newpage

\clearpage
\newpage

\end{document}